\documentclass[sigconf]{acmart}
\usepackage{bm}
\usepackage[subrefformat=parens]{subcaption}
\usepackage{booktabs}
\usepackage{extdash}

\setcopyright{acmcopyright}
\copyrightyear{2021}
\acmYear{2021}
\acmDOI{10.1145/3486622.3493947}

\acmConference[WI-IAT 2021]{WI-IAT 2021: The 20th IEEE/WIC/ACM International Conference on
Web Intelligence and Intelligent Agent Technology}{December 14--17, 2021}{Melbourne, Australia}
\acmBooktitle{WI-IAT 2021: The 20th IEEE/WIC/ACM International Conference on
Web Intelligence and Intelligent Agent Technology,
  December 14--17, 2021, Melbourne, Australia}
\acmISBN{978-1-4503-9115-3/21/12}

\begin{document}

\title{Analysis of Leading Communities Contributing to arXiv Information Distribution on Twitter}

\author{Kyosuke Shimada}
\orcid{0000-0001-5354-2020}
\affiliation{%
    \institution{Wakayama University}
    \streetaddress{930 Sakaedani}
    \city{Wakayama-shi}
    \state{Wakayama}
    \country{Japan}
    \postcode{525--0035}
}
\email{s216327@wakayama-u.ac.jp}

\author{Kazuhiro Kazama}
\affiliation{%
    \institution{Wakayama University}
    \streetaddress{930 Sakaedani}
    \city{Wakayama-shi}
    \state{Wakayama}
    \country{Japan}
}
\email{kazama@wakayama-u.ac.jp}

\author{Mitsuo Yoshida}
\affiliation{%
    \institution{University of Tsukuba}
    \streetaddress{3--29--1 Otsuka}
    \city{Bunkyo-ku}
    \state{Tokyo}
    \country{Japan}
}
\email{mitsuo@gssm.otsuka.tsukuba.ac.jp}

\author{Ikki Ohmukai}
\affiliation{%
    \institution{The University of Tokyo}
    \streetaddress{7--3--1 Hongo}
    \city{Bunkyo-ku}
    \state{Tokyo}
    \country{Japan}
}
\email{i2k@l.u-tokyo.ac.jp}

\author{Sho Sato}
\affiliation{%
    \institution{Doshisha University}
    \streetaddress{Karasuma-higashi-iru, Imadegawa-dori}
    \city{Kyoto-shi}
    \state{Kyoto}
    \country{Japan}
    \postcode{78229}
}
\email{min2fly@slis.doshisha.ac.jp}

\renewcommand{\shortauthors}{Shimada et al.}

\begin{abstract}
    To analyze the impact that arXiv is having on the world, in this paper we propose an arXiv information distribution model on Twitter, which has a three-layer structure: arXiv papers, information spreaders, and information collectors.
    First, we use the HITS algorithm to analyze the arXiv information diffusion network with users as nodes, which is created from three types of behavior on Twitter regarding arXiv papers: tweeting, retweeting, and liking.
    Next, we extract communities from the network of information spreaders with positive authority and hub degrees using the Louvain method,
    and analyze the relationship and roles of information spreaders in communities using research field, linguistic, and temporal characteristics.
    From our analysis using the tweet and arXiv datasets,
    we found that information about arXiv papers circulates on Twitter from information spreaders to information collectors, and that multiple communities of information spreaders are formed according to their research fields.
    It was also found that different communities were formed in the same research field, depending on the research or cultural background of the information spreaders.
    We were able to identify two types of key persons: information spreaders who lead the relevant field in the international community and information spreaders who bridge the regional and international communities using English and their native language.
    In addition, we found that it takes some time to gain trust as an information spreader.
\end{abstract}



\maketitle

\section{Introduction}

In recent years, preprint servers such as arXiv have been actively used for research information exchange and conference management, and it has been pointed out that Twitter plays a particularly important role in such academic information distribution
\cite{chiarelli_andrea_2019_3357727}.
It is assumed that the characteristics of information distribution on Twitter differ greatly from that of the conventional one-way information distribution using academic journals as a medium.
It is important to clarify its characteristics in order to understand contemporary academic information distribution.
For example, one of the possible characteristics is the role that Twitter users play in the distribution of preprints.
Expert users in the relevant research field discover important arXiv papers and introduce them on Twitter.
When users who read the tweets decide that the tweets are important and repeatedly like and retweet them, the information spreads widely.
In other words, users involved in arXiv information distribution 
play two types of roles: information spreaders and information collectors, 
and the strength of each aspect varies greatly depending on 
the characteristics of the user.

In this paper, we model arXiv information distribution on Twitter in three layers: arXiv papers, information spreaders, and information collectors, especially focusing on information spreaders, who we believe play a particularly important role.
First, we use the Hyperlink-Induced Topic Search (HITS) algorithm to analyze the arXiv information diffusion network with users as nodes, which is created from three types of behavior on Twitter regarding arXiv papers: tweeting, retweeting, and liking.
Next, we extract communities from the network of information spreaders with positive authority and hub degrees using the Louvain method, 
and analyze the relationship and roles of information spreaders in communities using research field, linguistic, and temporal characteristics.

\section{RELATED WORK}

\begin{sloppypar}
There is some relevant research available on information retrieval, the sharing of arXiv preprints, and social media.
\end{sloppypar}

A user survey by the arXiv team at Cornell University found that the impact of  Social Networking Services (SNS) was not evaluated because there were no SNS options in the answers. However, they were mainly looking for arXiv preprints from Google search, Google Scholar, and the actual arXiv homepage~\cite{rieger2016arxiv25}. 

It has been reported that even in SNS, especially on Twitter, 
the users who spread academic information are not necessarily academic, 
but many of them have backgrounds in social sciences and humanities~\cite{10.1371/journal.pone.0197265,10.1371/journal.pone.0175368}, 
and academic users follow the bots of each preprint server to obtain information, spread preprint information, 
and discuss it~\cite{chiarelli_andrea_2019_3357727}. 
In other words, Twitter plays an essential role in the distribution of academic information.

Regarding preprints that have changed with the development of the web, Boya et al.~\cite{mspreprint2021}
quantitatively showed the overall trends and impacts using the preprint datasets collected in the Microsoft Academic Graph (MAG) from 1991 to 2019~\cite{mag2015,shen-etal-2018-web}.
Preprint data of arXiv accounted for about 60\% of the overall sample.
In particular, most physics, mathematics, and computer science fields are posted on arXiv.
The authors noted that the number of preprints in computer science and biology has been increasing over the last ten years. Unlike other fields, international conferences are the main forum for presenting results in computer science, for example, and many preprints are submitted to conferences on machine learning.

Many of the studies on arXiv have evaluated data in specific fields, such as physics and mathematics~\cite{WANG2020101097,Gentil-Beccot:1214893}. 
However, Charles et al.~\cite{sutton2017popularity} and Jialiang et al.~\cite{howmanyPreprint} have conducted 
surveys limited to the computer science field~\cite{DBLP:journals/jcd/Halpern00}, which is growing particularly rapidly.
Charles et al. derived the percentage of papers published in arXiv by cross-referencing between DBLP and arXiv preprints and clarified the usage environment of arXiv in the computer science field.
Therein, the ratio of electronic editions was exceptionally high in theoretical computer science and machine learning. 
Meanwhile, though the usage rate is increasing in other fields, it still remains close to zero.

Jialiang et al. quantified how many preprints submitted to arXiv were eventually published in peer-reviewed journals.
The preprints posted in the computer science field of arXiv from 2008 to 2017 were investigated using Bidirectional Encoder Representations from Transformers (BERT), and the changes from the preprint version to the official publication were captured.

In addition, arXiv users may submit a preprint to arXiv and then publish the paper in a journal or international conference after peer review. 
Several studies have analyzed such usage patterns using academic information databases.

Larivi\`{e}re et al.~\cite{doi:10.1002/asi.23044arXivToDB} analyzed two data sources, arXiv and Web of Science (WoS). They found that about 64\% of the arXiv preprints between 1991 and 2011 were included in WoS, and 93\% of those were in mathematics, physics, earth science, and space science.
In particular, in astronomy, astrophysics, nuclear physics, and particle physics, most of the papers included in WoS were also submitted as preprints to arXiv.
In mathematics and physics, a high percentage also contributed arXiv preprints, though the percentage was low in certain sub-fields.

Shuai et al.~\cite{10.1371/journal.pone.0047523OnlineReaction} analyzed the relationship between the number of mentions on Twitter, the number of arXiv downloads, and the number of citations of papers for 4,606 preprints between October 2010 and May 2011.
As a result, it was clarified that the number of mentions on Twitter has a moderate correlation with the number of arXiv downloads several months after the preprint posting, as well as the number of early citations.
Many of the preprint subjects downloaded from arXiv and mentioned on Twitter were in the realms of astrophysics, high energy physics, and mathematics, with nearly 70\% of the papers peaking within five days of submission.

Based on the above, it is clear that studies on how users engage with academic information and on the relationship between preprints and peer-reviewed papers are available; however, such research has not considered the dual role of users in collecting and disseminating academic information on social media.
Thus, we believe it is necessary to investigate such users, who contribute to the spread of arXiv preprints, and examine their duality.

\section{Analysis methods for arXiv information distribution}

\subsection{Three-layer model of arXiv information distribution}
\begin{figure}[tb]
	\centering
	  \includegraphics[width=6.1cm]{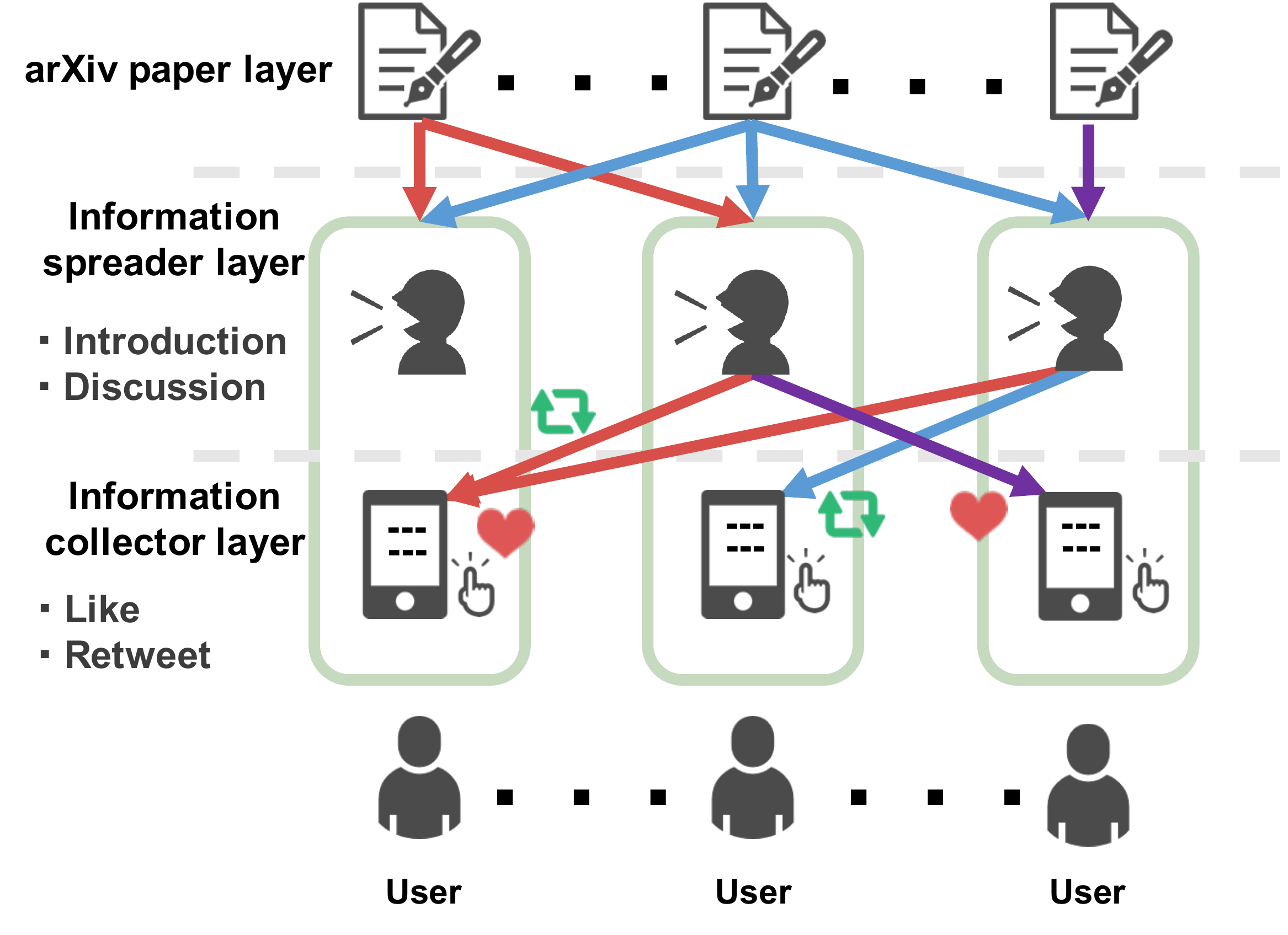}
	  \caption{Three-layer model of arXiv information distribution}
	  \label{fig:three-layer-model}
\end{figure}

In this paper, arXiv information distribution on Twitter is presented as a three-layer model, as shown in Figure~\ref{fig:three-layer-model}, instead of detecting or estimating the diffusion path of the arXiv paper's information on a graph of users or tweets.
The first layer comprises arXiv papers.
The reason we call them ``arXiv papers'' instead of ``arXiv preprints'' is because we do not use other databases in order to narrow the scope to preprints only.
The second layer consists of information spreaders, who spread information by tweeting or retweeting the URL of arXiv articles.
The third layer is made up of information collectors, who retweet or like tweets only if they find the information to be valuable.
In this model, we assume that the same user can have two different roles for arXiv papers: information spreader and information collector.

Using this model, we focus on the information spreaders, who contribute to arXiv information distribution in particular, and analyze them in terms of the importance of the user, the community to which the user belongs, and the characteristics of the users or communities.

\subsection{Importance of information spreaders and information collectors}\label{sec:hits}
On Twitter, information spreaders who are retweeted or liked by many information collectors are considered important and reliable, while information collectors who retweet or like many tweets of important information spreaders are considered trustworthy.
Thereby, we created an information diffusion network with users as nodes by combining the information diffusion of arXiv papers on Twitter.
We used the authority and the hub weight, which are calculated by Kleinberg's HITS algorithm~\cite{Kleinberg:1999:ASH:324133.324140}, representing the importance of an information spreader and an information collector, respectively.

Suppose that a user $u_i (i=1, \ldots{}, N)$ performs two roles as an information spreader $u^s_i$ and an information collector $u^c_i$.
When an information collector $u^c_i$ retweets or likes the tweet of an information spreader $u^s_j$, the information is considered to have propagated from $u^s_j$ to $u^c_i$.
The information diffusion network of the user can be represented by an adjacency matrix $\bm{D}$ with $N$ rows and $N$ columns whose elements are the diffusion $d_{i,j}$.
The authority weight $a_i$ and hub weight $h_i$ of user $u_i$ are obtained by the following procedure:

\begin{sloppypar}
\begin{enumerate}
    \item Initialize each element of an authority vector \hbox{$\bm{a}=(a_1, \ldots, a_N)^\top$}, which represents the importance of each user as an information spreader, and a hub vector \hbox{$\bm{h}=(h_1, \ldots, h_N)^\top$}, which represents the importance of each user as an information collector, with 1.
    
    \item Using an adjacency matrix $\bm{D}$, repeat the normalization after applying Equations~\ref{eq:authority} and \ref{eq:hub} until the values of $\bm{a}$ and $\bm{h}$ converge.
          \begin{eqnarray}\label{eq:hits_equation}
              \bm{a}&=&\bm{D}^\top \bm{h}\label{eq:authority}\\
              \bm{h}&=&\bm{D} \bm{a}\label{eq:hub}
          \end{eqnarray}

    \item From this result, create a pair $(a_i, h_i)$ of authority weight and hub weight designations for user $u_i$.
\end{enumerate}
\end{sloppypar}

Degree centrality, such as indegree and outdegree, is easily affected by the size of an information spreader community and tends to be evaluated higher for nodes in large communities.
However, the authority and hub weight degrees tend to be higher when there is a bipartite graph structure of information spreaders and information collectors in the community, making them more suitable for comparison and analysis across multiple communities than degree centrality.

\subsection{Information spreader community}\label{sec:louvain}
We created an information spreader network focusing on information spreaders who contribute to arXiv information distribution and extracted similar information spreader communities from the viewpoint of information collectors using the following procedure:

\begin{enumerate}
    \item Create a set $U^c_i=\{u^c_j | d_{j, i} = 1\}$ of information collectors who retweeted and liked information spreader $u^s_i$ satisfying $a_i > 0 \wedge h_i > 0$.

    \item If the Szymkiewicz\Hyphdash Simpson coefficient $s(U^c_i, U^c_j)$ of the information spreaders $u^s_i$ and $u^s_j$ is greater than or equal to the threshold $T$, the edge $e_{i,j}$ is stretched and the information spreader network $G^s$ is created.

    \item Split $G^s$ into communities $C^s_1, \ldots, C^s_K$ using the Louvain method~\cite{Blondel_2008}.
\end{enumerate}

The reason for limiting the analysis to users with positive authority and hub weights is that users with zero authority weight are not involved in information diffusion, and users with positive authority weights but zero hub weights are either bots or accounts dedicated to information diffusion, so they are excluded from the analysis.

In addition, we used betweenness centrality~\cite{freeman1977} in the information spreader network to identify important information spreaders that interconnect multiple information spreader communities.

\subsection{Characteristics of information spreader communities}
We analyzed information spreader communities from three perspectives: the research field characteristics of users, linguistic characteristics, and temporal characteristics.

\subsubsection{Research field characteristics of users}
Assuming that users' expertise is manifested in the categories of arXiv papers they spread or collect information from, we analyzed the academic trends of information spreader communities using the categories of arXiv papers.

The arXiv papers are stored in 11 different archives, such as Computer Science (e.g., cs), and the categories are represented as a string with the archive name as 
the major category, with sub-categories of research fields added on after periods (e.g., cs.AI).
In addition, physics is divided into several categories.
Therefore, gr-qc (General Relativity and Quantum Cosmology), nlin (Nonlinear Sciences), nucl (Nuclear Theory, Nuclear Experiment), and quant-ph (Quantum Physics) are collectively denoted as physics*, as in ``arXiv submission rate statistics''\footnote{\url{https://arxiv.org/help/stats/2019_by_area/}}.

Since information spreaders are also information collectors, we distinguished between the categories of arXiv papers that spread information and those that collected information by calling them \emph{information spread category} and \emph{information collection category}, respectively.
For example, it is thought that the information spread category represents the user's current expertise, while the information collection category represents the fields that the user would like to refer to in the future.

In arXiv, multiple categories can be assigned to a single paper;  however, we use only the first assigned category in this analysis. We call this the main category of the arXiv paper.
In addition, for each user, the main categories of the articles hat were used to spread or collect information are obtained, and the category with the largest number is used as the user's information spread category or information collection category.

\subsubsection{Linguistic characteristics of users}
Since the cultural background of users is reflected in the language they use, we analyzed the cultural trends of information spreader communities using the language information about users in each community.

On Twitter, people tend to use their native language or the language of their organization for daily communication such as tweeting and replying.
This is called \emph{communication language} (CL).
On the other hand, users may use a language different from their mother tongue, such as English for example, when presenting international academic papers.
This is called \emph{profile language} (PL).
The linguistic characteristics of a user or a community on Twitter are defined using these two types of linguistic identifiers.

For the communication language, we use the lang field obtained using the Twitter API.
The profile language is determined using Python's langdetect\footnote{\url{https://pypi.org/project/langdetect/}} library after URL removal from the profile text.
In both cases, the language is represented by a two-letter lowercase alphabet as specified in ISO 639-1.
However, if the Twitter API or langdetect cannot determine the language, it is expressed as ``UD'' (undecided).

\subsubsection{Temporal characteristics of users}
Since the degree of influence that a user has on Twitter is considered to be determined by the period that the user was actively speaking,
we analyzed the activity of the information spreader community using the \emph{mention period} of each user.
The mention period is the number of days between the first and last tweets mentioning any arXiv papers.
However, this is limited to cases where there are two or more such tweets.

\section{Analysis}

\subsection{Datasets}
We used the Twiter API to collect tweets about arXiv papers between March 21, 2007 and January 18, 2020.
The compressed URLs in the tweets were decompressed and used as the \emph{mention dataset} for arXiv papers.
The details of the mention dataset are shown in Table~\ref{tab:mention_analysis}.
Notably, there is a restriction in the Twitter API that means only a maximum of 100 retweets and likes can be obtained for a tweet.
However, since there were only 5,600 cases violating this limit for likes and 1,449 cases for retweets, it is not considered to be a major problem.

\begin{table}[t]
    \centering
    \footnotesize
    \caption{Details of the mention dataset}
    \begin{tabular}{lrl}
        \toprule
        Number of mentioned papers & 981,865\\
        Total number of mention tweets                & 3,088,669 & (100\%)\\
        Number of liked tweets     & 797,294 & (25.8\%)\\
        Number of retweeted tweets & 446,652 & (14.5\%)\\
        Total number of users                 & 586,999 & (100\%)\\
        Number of users mentioned           & 118,743 & (20.2\%)                 \\
        \bottomrule
    \end{tabular}
    \label{tab:mention_analysis}
\end{table}

The details of the authority and hub weights of the users are shown in Table~\ref{tab:result_n}.
The number of users with positive authority is 11.0\% of the total, indicating that only a small portion of the users involved in arXiv information distribution are information spreaders.
Furthermore, there are 3.5\% of users with positive authority but zero hub degree, because they send out information on arXiv papers but do not receive information from others.
These were removed from the scope of the information spreader network.

The number of nodes and edges in the information spreader network were 5,497 and 20,453, respectively, based on the Szymkiewicz\Hyphdash Simpson coefficient threshold of 0.5.
In addition, the Louvain method was applied to the information spreader network, resulting in 169 communities.
Its largest connected component had 5,034 nodes (91.6\%) and 20,119 edges (98.4\%), and contained 26 communities.

\begin{table}[t]
    \centering
    \footnotesize
    \caption{Details of authority and hub weights}
    \begin{tabular}{lrl}
            \toprule
            Total number of users & 586,999 & (100\%)\\
            $a_i > 0$ & 64,490 & (11.0\%) \\
            $h_i > 0$ & 566,367 & (96.5\%) \\
            $a_i > 0 \wedge h_i > 0$ & 43,858 & (7.5\%) \\
            $a_i > 0 \wedge h_i = 0$ & 20,632 & (3.5\%)  \\
            $a_i = 0 \wedge h_i > 0$ & 522,509 & (89.0\%) \\
            \bottomrule
        \end{tabular}
    \label{tab:result_n}
\end{table}

In addition, the bibliographic information of 1,645,129 arXiv papers, which were collected from arXiv using OAI-PMH\footnote{\url{http://www.openarchives.org/OAI/openarchivesprotocol.html}}
on February 26, 2020, was used as the \emph{arXiv metadata set}.
This bibliographic information includes 
paper ID,
author(s), 
publish date,
updated date,
title,
category (main or secondary, multiple available),
abstract, comments, DOI, Journal-ref, Report number, ACM-class, and MSC-class.

\subsection{Analysis of time-series variation in the categories of mentioned papers}
We analyzed the time-series variation in the categories of arXiv papers mentioned on Twitter.

\begin{figure}[tb]
    \centering
    \includegraphics[width=6.5cm]{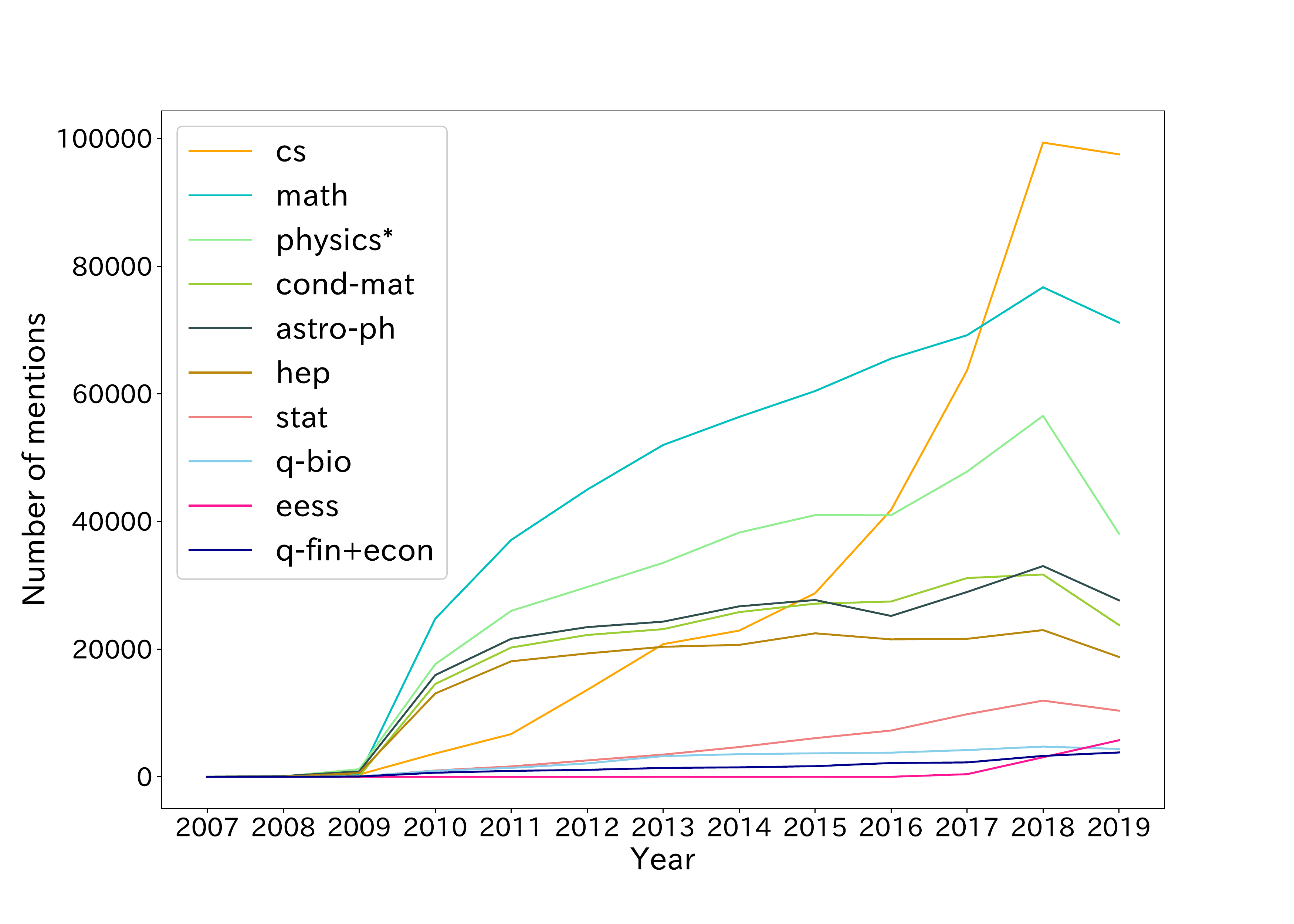}
    \caption{Time-series variation in categories of mentioned papers}
    \label{fig:category_mention_change2}
\end{figure}

First, we show the time-series variation in the number of mentioned papers by the category of arXiv papers in Figure~\ref{fig:category_mention_change2}.
The horizontal axis represents the year, and the vertical axis represents the number of mentions of arXiv papers in each category for that year.
It can be seen that the number of mentions of arXiv papers has increased rapidly in several fields since 2009.
A closer look reveals that immediately after 2009, the fields of mathematics (math) and physics (physics*) were frequently mentioned, but after 2011, the mentions of the field of computer science (cs) rapidly increased, and after 2018, it became the most mentioned field.
Comparing with the number of submitted papers in ``arXiv submission rate statistic''\footnote{\url{https://arxiv.org/help/stats/2019_by_area/}}, we can see that the trend of the increase in the number of submitted papers and the number of mentions is basically similar, but in the field of computer science, the increase in the number of mentions is much larger than the number of submitted papers.

\begin{figure}[tb]
    \centering
    \includegraphics[width=6.5cm]{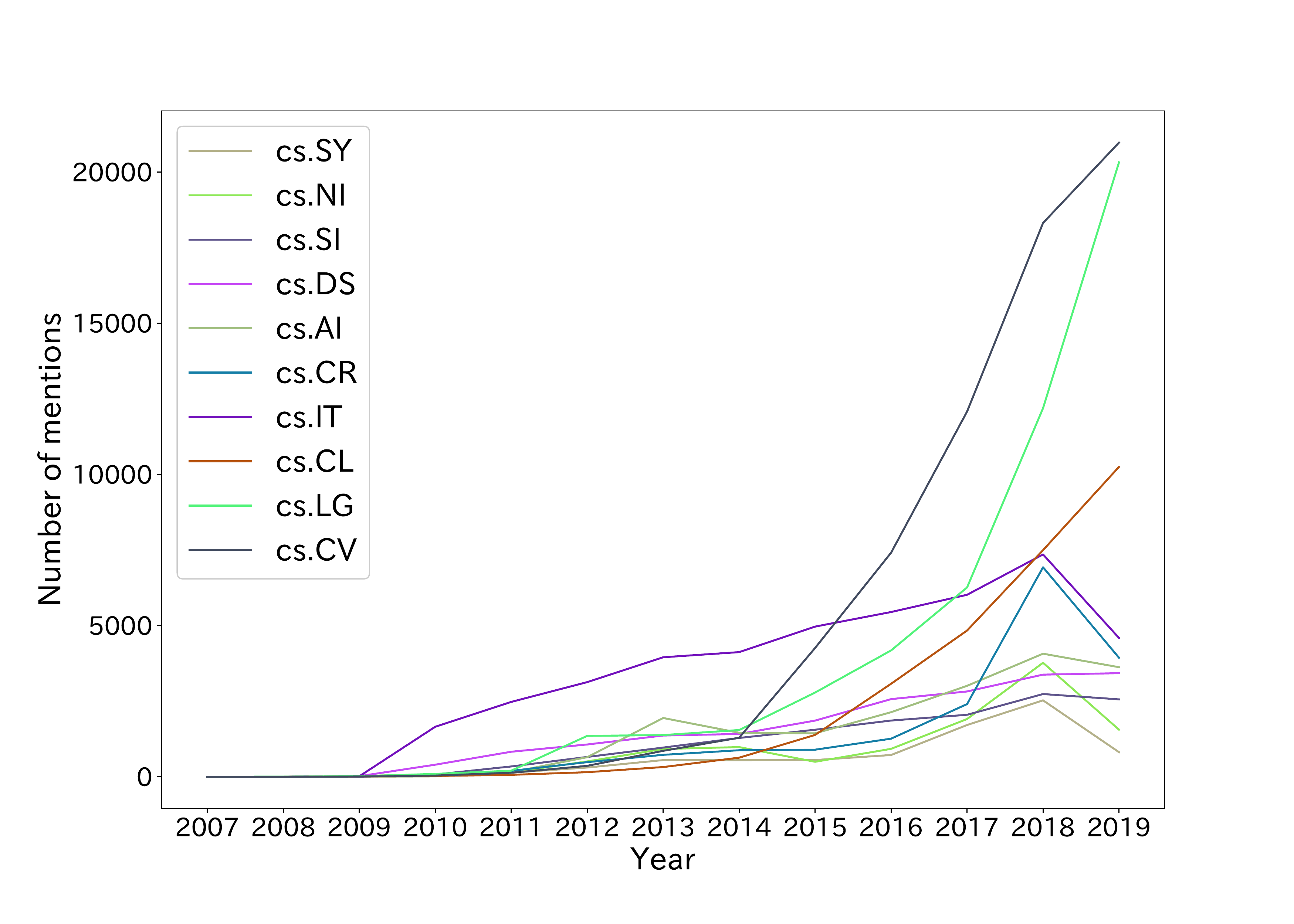}
    \caption{Time-series variation in sub-categories of mentioned papers in Computer Science}
    \label{fig:category_mention_change3}
\end{figure}

Next, we show the time-series change of detailed categories in computer science, where the number of mentions has been increasing, especially in recent years, in Figure~\ref{fig:category_mention_change3}.
The horizontal axis represents the year and the vertical axis represents the number of mentions of arXiv papers in each sub-category for that year.
The increase in the number of mentions in 2009 was mainly due to data structures and algorithms (cs.DS).
However, since 2014, the number of mentions of image recognition (cs.CV) and machine learning (cs.ML) have increased dramatically.
Since 2015, the number of mentions of natural language processing (cs.CL) have also increased rapidly.

From the above results, it can be seen that arXiv was utilized by researchers in mathematics and physics in the early stages.
However, recently, due to the rapid development of machine learning and deep learning, the use of arXiv by researchers in the field of computer science has increased significantly.
One reason for this may be that in computer science, there is a tendency to place a higher value on international conferences than on academic journals as a place to present papers~\cite{ConferencesVS}; 
the increasing use of preprint servers and Twitter in international conferences could be another major factor.

\subsection{Research field characteristics analysis of information spreader communities}
\begin{figure}[tb]
    \centering
    \includegraphics[width=7cm]{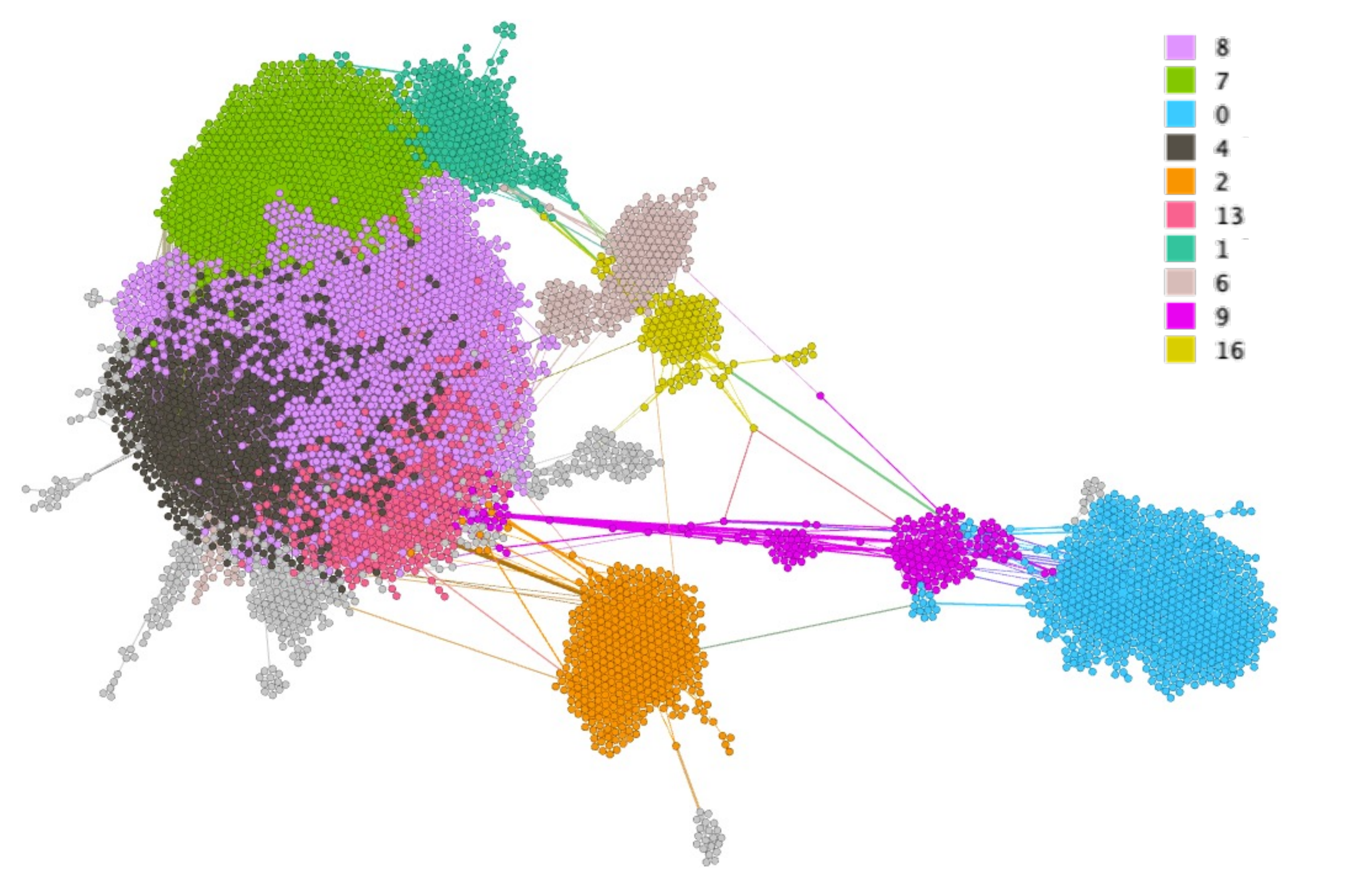}
    \caption{Visualization result of the information spreader network}
    \label{fig:overview}
\end{figure}

We show the visualization result of the community structure extracted from the maximum connected component of the information spreader network in Figure~\ref{fig:overview}.
After laying out nodes with Gephi's ForceAtlas2 algorithm, the node size was varied according to its authority weight and the edge thickness according to the Szymkiewicz\Hyphdash Simpson coefficient. The top 10 communities by the number of users are assigned different colors and the 15 remaining smaller communities are labeled by the same color.

To investigate the research field characteristics of information spreader communities, the categories of information spreaders and information gatherers and the number of users in the top three categories of the top 10 communities are shown in Table~\ref{tab:comcategory}.
The information spread categories refer to the category of arXiv papers mentioned by the user. The information collection categories constitute the category of arXiv papers that have been liked or retweeted. The top three categories and the number of users are shown, respectively.

From information spread categories, we can classify information spreader communities into two major research fields: physics and machine learning.
The dense communities 4, 7, 8, and 13 on the upper left of the visualization result are machine learning communities, and the dispersed communities 0, 2, 1, and 6 on the right are physics communities.
It can be seen that the information spreaders of the same research field can be divided into several communities.
Furthermore, when we compare information spread categories with information collect categories, they show essentially the same tendency.
However, in communities 8, 7, and 4, the number of users of machine learning (cs.LG) is significantly larger in the information collect category than in the information spread category.
This suggests that machine learning information is particularly sought after in these communities.

\begin{table*}[t]
    \centering
    \footnotesize
    \caption{Information spread and collect categories by community}
    \label{tab:comcategory}
    \begin{tabular}{rrll}
        \toprule
        CN & \multicolumn{1}{l}{\#Users}        & Information spread categories & Information collect categories\\
        \midrule
        8 & 1414 & cs.LG: 651, cs.CV: 249, cs.CL: 156             & cs.LG: 836, cs.CV: 175, cs.CL: 154\\
        7 & 685  & cs.LG: 206, cs.CV: 174, cs.CL: 62             & cs.LG: 310, cs.CV: 160, stat.ML: 54\\
        0 & 628  & astro-ph.EP: 196, astro-ph.GA: 165, astro-ph.HE: 55& astro-ph.EP: 211, astro-ph.GA: 198, astro-ph.HE: 61\\
        4 & 606  & cs.LG: 311, cs.CV: 99, stat.ML: 71              & cs.LG: 392,  cs.CV: 77, stat.ML: 54\\
        2 & 377  & physics.soc-ph: 177,  cs.SI: 47, q-bio.NC: 42    & physics.soc-ph: 226,  cs.SI: 38, q-bio.NC: 37\\
        13 & 314 & cs.CL: 255, cs.LG: 32, stat.ML: 6             & cs.CL: 260, cs.LG: 40, stat.AP: 2\\
        1 & 256  & hep-th: 43,  quant-ph: 29, hep-ph: 11           & hep-th: 46, quant-ph: 33, cond-mat.str-el: 17\\
        6 & 245  & quant-ph: 115, physics.chem-ph: 30, cs.LG: 20 & quant-ph: 141, physics.chem-ph: 35, cs.LG: 26\\
        9 & 184  & hep-ph: 52, hep-ex: 46, astro-ph.CO: 18             & hep-ex: 55, hep-ph: 52, astro-ph.CO: 16\\
        16 & 119 & math.CT: 23, cs.LO: 20, cs.PL: 16             & math.CT: 35, cs.PL: 20, cs.DC: 8\\
        \bottomrule
    \end{tabular}
\end{table*}

\subsection{Linguistic characteristics analysis of information spreader communities}
Next, we analyzed the linguistic characteristics of information spreader communities.
Table~\ref{tab:comlang} shows the communication languages and profile languages of each community in descending order of the top three number of users.
UD in this table represents the users whose language could not be identified.

The number one language in most communities is English.
However, it is Japanese in communities 7 and 1.
In the machine learning communities 8, 7, 4, and 13, 
Japanese is the main language used in the second-largest community 7, English is the main language used in the other communities, and the research fields of all communities are very similar.
In contrast, in the physics communities 0, 2, 1, and 6, Japanese is the main language used in community 1, while English is the main language used in the other communities, and the research fields of each community are different.

This result indicates that the arXiv paper information is mainly diffused internationally using English.
Even in communities 7 and 1, where Japanese is the main language, English is the second most commonly used language.
However, since Japanese is in a very different language family from English, it can be assumed that Japanese is used for communication among Japanese researchers or developers who have many opportunities to interact with each other daily.

Furthermore, in communities 7 and 1, there are more English users and fewer Japanese users in profile languages than in communication languages.
This may be because they write their profiles in English for international academic activities, but communicate in their native language for regional communication and contributions.
In a research field such as machine learning, which has had a large impact on the world, it can be assumed that in addition to an international community that uses English as a common language for research activities, a regional community has emerged that communicates using its native language while focusing on the same paper.
When we examined the non-maximum connected component, we found community 105, whose main communication and profile language is Korean, and community 83, whose main communication language is Japanese; however, these are small, isolated communities with six and two users, respectively.

\begin{table}[t]
    \centering
    \footnotesize
    \caption{Communication languages and profile languages of each community}
    \label{tab:comlang}
    \begin{tabular}{rrll}
        \toprule
        CN & \multicolumn{1}{l}{\#Users}        & CL & PL\\
        \midrule
        8  & 1414 & en:1341, UD:19, fr:2                  & en:1244, UD:63, de:21 \\
        7  & 685 & ja:506, en:121, UD:22                 & ja:346, en:218, UD:26 \\
        0  & 628 & en:583, UD:5, es:4                    & en:546, UD:32, it:8   \\
        4  & 606 & en:568, UD:8, pt:4                    & en:528, UD:33, de:12  \\
         2  & 377 & en:357, tl:2, ca:1                     & en:326, UD:19, es:4   \\
        13 & 314 & en:302, UD:6, ja:1                    & en:276, UD:18, de:10  \\
       1  & 256 & ja:182, en:42, UD:10                  & ja:154, en:62, ko:11   \\
        6  & 245 & en:233, UD:2, ja:2                    & en:213, UD:17, fr:3   \\
        9  & 184  & en:165, es:9, UD:3                    & en:154, es:11, UD:10  \\
        16  & 119 & en:111, ro:2, es:2                     & en:91, UD:14, de:2   \\
        \bottomrule
    \end{tabular}
\end{table}

\subsection{Analysis of key people in information spreader communities}
We analyzed key people in information spreader communities using authority weight and betweenness centrality.

First, we analyzed the top 20 users in authority weight.
The ranking of the authority weight, screen names, community numbers (CN), communication languages (CL), profile languages (PL), authority weights ($a_i$), and hub weights ($h_i$) are shown in Table~\ref{tab:hits}.
The numbers in parentheses in this table are the rankings in hub weight.
Users with hub weight 0 are not included in the information spreader network, so they are not assigned community numbers.
Table~\ref{tab:hits} includes Twitter accounts from prominent researchers such as Miles Brundage (@Miles\_Brundage), who is a member of OpenAI, and Ian Goodfellow (@goodfellow\_ian), who proposed GANs; prominent companies such as DeepMind (@DeepMind), which developed AlphaGo; and bots such as @arxiv\_org and @StatMLPapers.
However, their hub weights are not necessarily high, especially for bots, which have a hub weight of 0.
When we examined which communities the top 20 information spreaders belonged to, we found that most of them belonged to communities 8 or 4.

\begin{table}[t]
    \centering
    \footnotesize
    \caption{Rankings of the top 20 information spreaders by authority weight}
    \label{tab:hits}
    \begin{tabular}{rlllrll}
        \toprule
         & screen name          & CN & CL & PL & $a_i$ & $h_i$\\
        \midrule
        1  & @hardmaru          &  8 &  en &  en & 0.007356 & 0.000295 (25) \\
        2  & @Miles\_Brundage   & 8 & en & en & 0.006711    & 0.000219 (79) \\
        3  & @arxiv\_org        &  - & en & en &0.005886    & 0.000000 (574486)  \\
        4  & @DeepMind          & 8 & en & en & 0.005608    & 0.000004 (50110)  \\
        5  & @StatMLPapers      & - & en & en &0.004696     & 0.000000 (575126) \\
        6  & @goodfellow\_ian   & 4 & en & en & 0.003937    & 0.000095 (995)  \\
        7  & @alexjc            & 4 & en & en & 0.003858    & 0.000094 (1013) \\
        8  & @quantombone       & 8 & en & en &0.003768     & 0.000068 (1860) \\
        9  & @Reza\_Zadeh       & 4  & en & en &0.003619    & 0.000018 (10989)  \\
        10 & @evolvingstuff     & 4 & en & en &0.003618     & 0.000193 (132) \\
        11 & @ml\_review        & 4 & en & en &0.003578  & 0.000048 (3264) \\
        12 & @karpathy          & 4 & en & en & 0.003373  & 0.000063 (2099)  \\
        13 & @samim             & 4 & en & en & 0.003348  & 0.000138 (392)  \\
        14 & @fchollet          & 4 & en & en & 0.003342  & 0.000073 (1615) \\
        15 & @rsalakhu          & 4 & en & en & 0.003288  & 0.000051 (3006) \\
        16 & @weballergy            & 4 & en & en & 0.003281  & 0.000208 (108) \\
        17 & @hillbig        & 7& ja & en & 0.003160  & 0.000002 (90321) \\
        18 & @dennybritz              & 4 & en & en & 0.003028  & 0.000075 (1577)  \\
        19 & @OriolVinyalsML          & 4 & en & en & 0.003011  & 0.000035 (5045)  \\
        20 & @NandoDF         & 4 & en & en & 0.002835  & 0.000163 (248)  \\
        \bottomrule
        \end{tabular}
\end{table}

Next, we analyzed the top 20 users in betweenness centrality to find out which users act as bridges between communities.
The rankings of the authority weight, screen names, community numbers (CN), communication languages (CL), profile languages (PL), and betweenness centrality ($b_i$) are shown in Table~\ref{tab:baikair}.

In betweenness centrality, the top users changed significantly from authority weight.
Many users in communities 8 and 4 decreased in rank, while users in community 7 increased in rank.
The number of users belonging to communities other than communities 8, 4, and 7 also increased.

In particular, when we focus on regional community 7 of machine learning, Daisuke Okanohara (@hillbig), the COO/representative director of machine learning start-ups, ranks 17th in authority weight and 4th in betweenness centrality, and Yuta Kashino (@yutakashino), an entrepreneur, ranks 68th in authority weight and 20th in betweenness centrality, which is a remarkable improvement.
These users are spreading information to their community by introducing the arXiv paper in Japanese.
Their profile language is English and their communication language is Japanese,
 indicating that they are key persons with special roles, bridging the regional and international communities.

\begin{table}[t]
    \centering
    \footnotesize
    \caption{Rankings of the top 20 information spreaders by betweenness centrality}
    \label{tab:baikair}
    \begin{tabular}{rlllrl}
        \toprule
         & screen name          & CN & CL & PL & $b_i$ \\
        \midrule
        1 & @hardmaru           & 8 & en    & en    & 0.249 (1)\\
        2 & @Miles\_Brundage    & 8 & en    & en    & 0.226 (2)\\
        3 & @KyleCranmer        & 9 & en    & en    & 0.171 (72)\\
        4 & @hillbig            & 7 & ja    & en    & 0.142 (17)\\
        5 & @gfbertone          & 9 & en    & en    & 0.128 (1280)\\
        6 & @IntelligenceTV     & 4 & en    & en    & 0.111 (3629)\\
        7 & @alexvespi          & 2 & en    & en    & 0.091 (87)\\
        8 & @cloud149           & 0 & en    & en    & 0.088 (24873)\\
        9 & @pietrovischia      & 9 & en    & en    & 0.048 (19171)\\
        10 & @FuzzyDarkMatter   & 0 & en    & en    & 0.045 (9644)\\
        11 & @kaustuvdatta7C9   & 9 & en    & en    & 0.038924  (8586)\\
        12 & @Holger\_Schulz    & 9 & en    & UD    & 0.0354691 (6368)\\
        13 & @phi\_nate         & 9 & en    & en    & 0.0319739 (7194)\\
        14 & @quantombone       & 8 & en    & en    & 0.0295081 (8)\\
        15 & @DeepMind          & 8 & en    & en    & 0.0270675 (4)\\
        16 & @ragreens          & 10 & en   & UD    & 0.0259773 (9139)\\
        17 & @math3ma           & 16 & en   & en    & 0.025891  (402)\\
        18 & @rgmelko           & 6 & en    & en    & 0.0237524 (651)\\
        19 & @TomBreloff        & 4 & en    & en    & 0.0235911 (3912)\\
        20 & @yutakashino       & 7 & ja    & en    & 0.0234972 (68)\\
        \bottomrule
    \end{tabular}
\end{table}

\begin{figure*}[tb]
    \centering
    \subcaptionbox{\label{fig:authclear}authority weight}
    {
        \includegraphics[width=6.5cm]{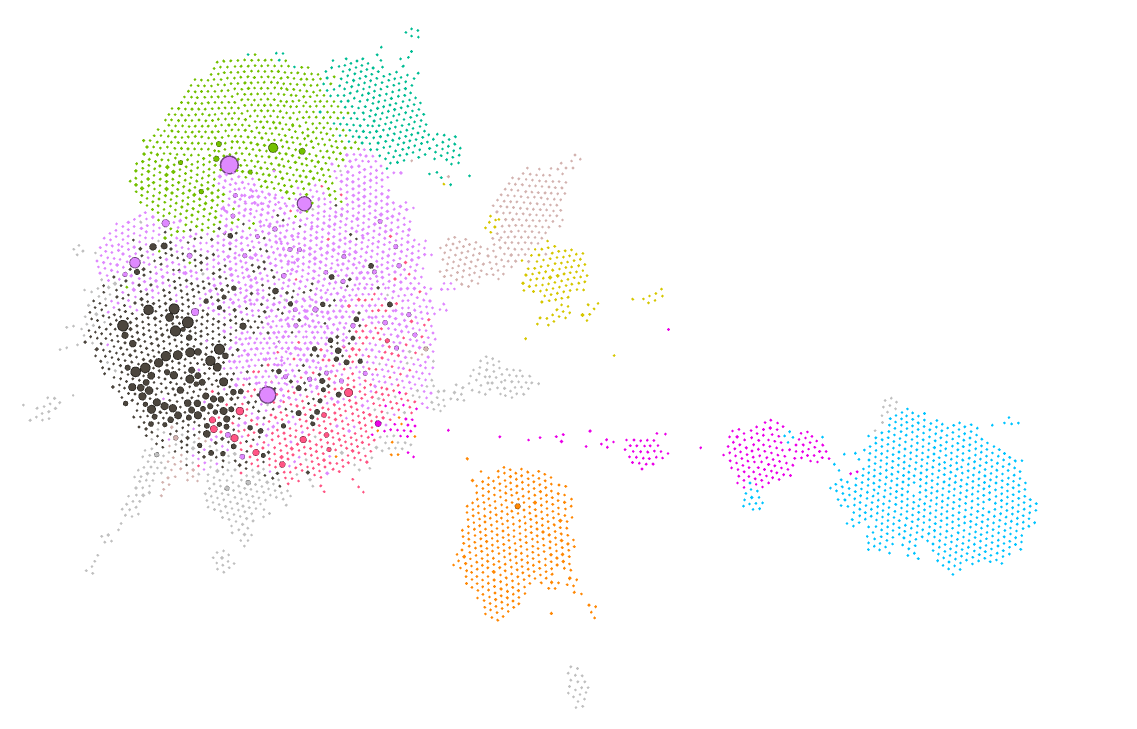}
    }
    \subcaptionbox{\label{fig:bcclear}betweenness centrality}
    {
        \includegraphics[width=6.5cm]{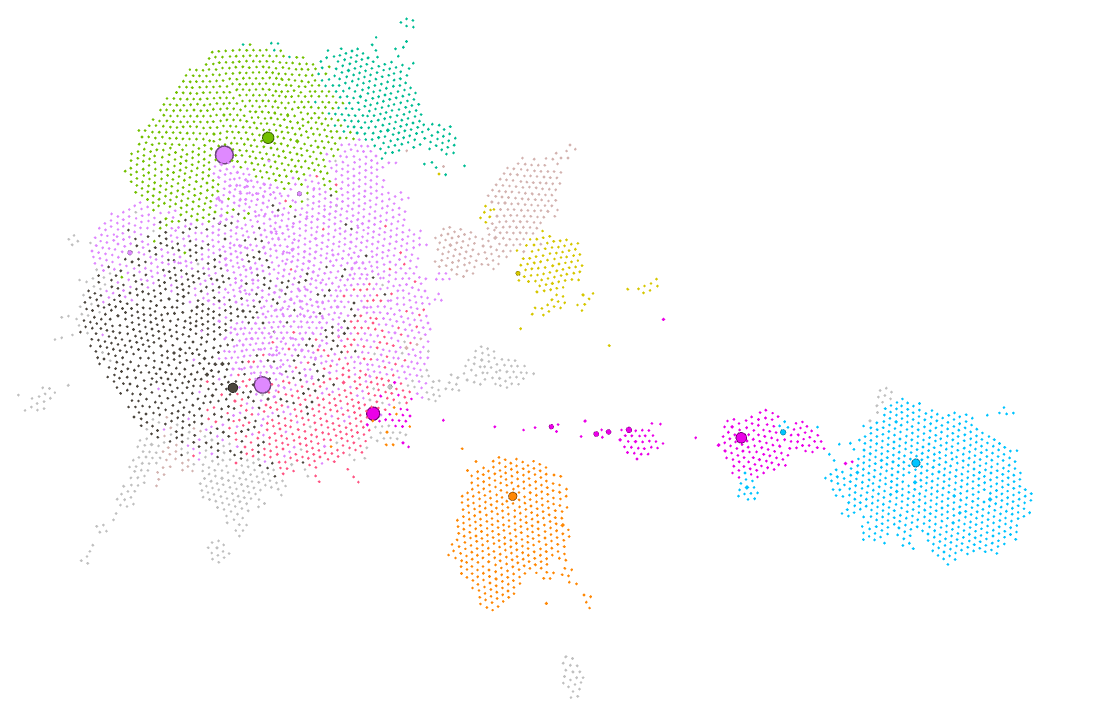}
    }

    \caption{Visualization results of key people in information spreader communities}
    \label{fig:nodeview}
\end{figure*}

Next, we visualized the positions of key people with high authority weight and betweenness centrality in the information spreader network.
The results of the visualization are shown in Figure~\ref{fig:nodeview}.
Only the nodes of the same information spreader network as Figure~\ref{fig:overview} are drawn in the same arrangement, with the size of nodes increasing for higher authority weight or betweenness centrality, and the transparency of nodes increasing for lower authority weight or betweenness centrality.

Figure~\ref{fig:overview}\subref{fig:authclear} shows that the information spreaders with high authority weight exist in communities 8, 7, and 4, which are relatively large.
In contrast, Figure~\ref{fig:overview}\subref{fig:bcclear} shows that the information spreaders with high betweenness centrality are distributed in more communities, although they are less likely to be in a single community.

\subsection{Temporal characteristics analysis of information spreaders}
Finally, we analyzed the temporal characteristics of the behavior of information spreaders.

First, we show the mean, maximum, median, and standard deviation ($\sigma$) of the mention period for each community in Table~\ref{tab:comtime}.
The results show that most of the communities tend to have long mention periods.

\begin{table}[t]
	\centering
	\footnotesize
	\caption{Statistics of mention periods in each community}
	\label{tab:comtime}
	\begin{tabular}{rrrrr}
		\toprule
		CN & \multicolumn{1}{l}{Mean} & \multicolumn{1}{l}{Maximum} & \multicolumn{1}{l}{Median} & \multicolumn{1}{l}{$\sigma$} \\
		\midrule
		8            & 734.5  & 4201.0 & 565.0  & 710.6\\
		7            & 1038.5 & 4306.0 & 810.5  & 914.5\\
		0            & 1240.9 & 4231.0 & 1068.0 & 982.5\\
		4            & 896.4  & 3848.0 & 776.0  & 803.2\\
		2            & 1286.6 & 3967.0 & 1126.5 & 1006.9\\
		13           & 537.7  & 3279.0 & 365.5  & 578.9\\
		1            & 1634.0 & 4163.0 & 1340.0 & 1089.1\\
		6            & 884.7  & 3835.0 & 654.0  & 898.7\\
		9            & 1148.4 & 3868.0 & 815.0  & 1008.7\\
		16           & 1120.5 & 4050.0 & 803.5  & 993.1\\
		\bottomrule
	\end{tabular}
\end{table}

\begin{figure}[tb]
	\centering
	\includegraphics[width=6.5cm]{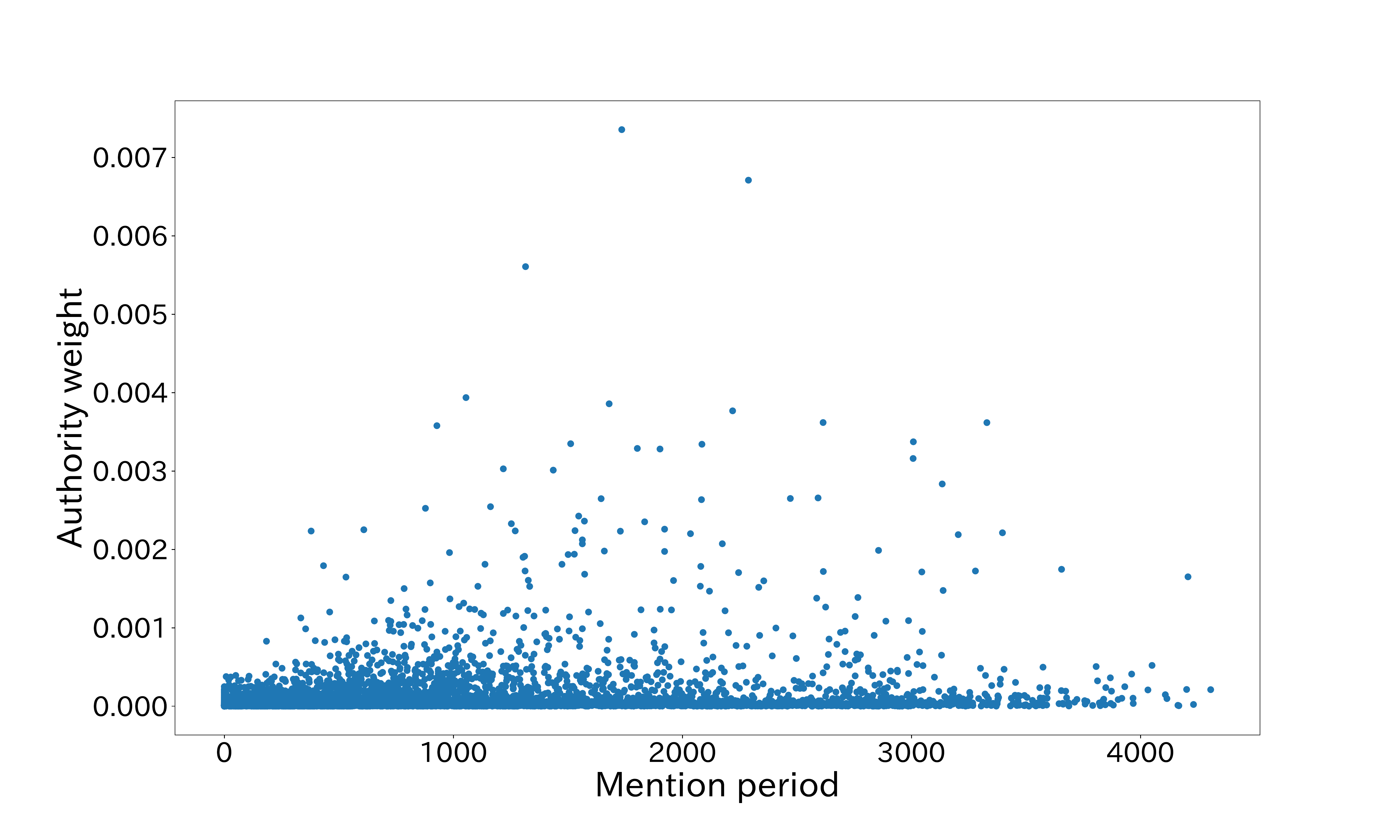}
	\caption{Distribution of mention periods and authority weights}
	\label{fig:mention_authority}
\end{figure}

In addition, we show the relationship between mention periods and authority weights in Figure~\ref{fig:mention_authority}.
From this result, it can be seen that users with a long mention period do not necessarily have high authority weights, but the authority weight does not increase unless the mention period is at least somewhat long.
In other words, 
relatively long-term activities by a key person in the arXiv information distribution network are correlated with a high authority weight.

\begin{figure}[tb]
	\centering
	\includegraphics[width=6.5cm]{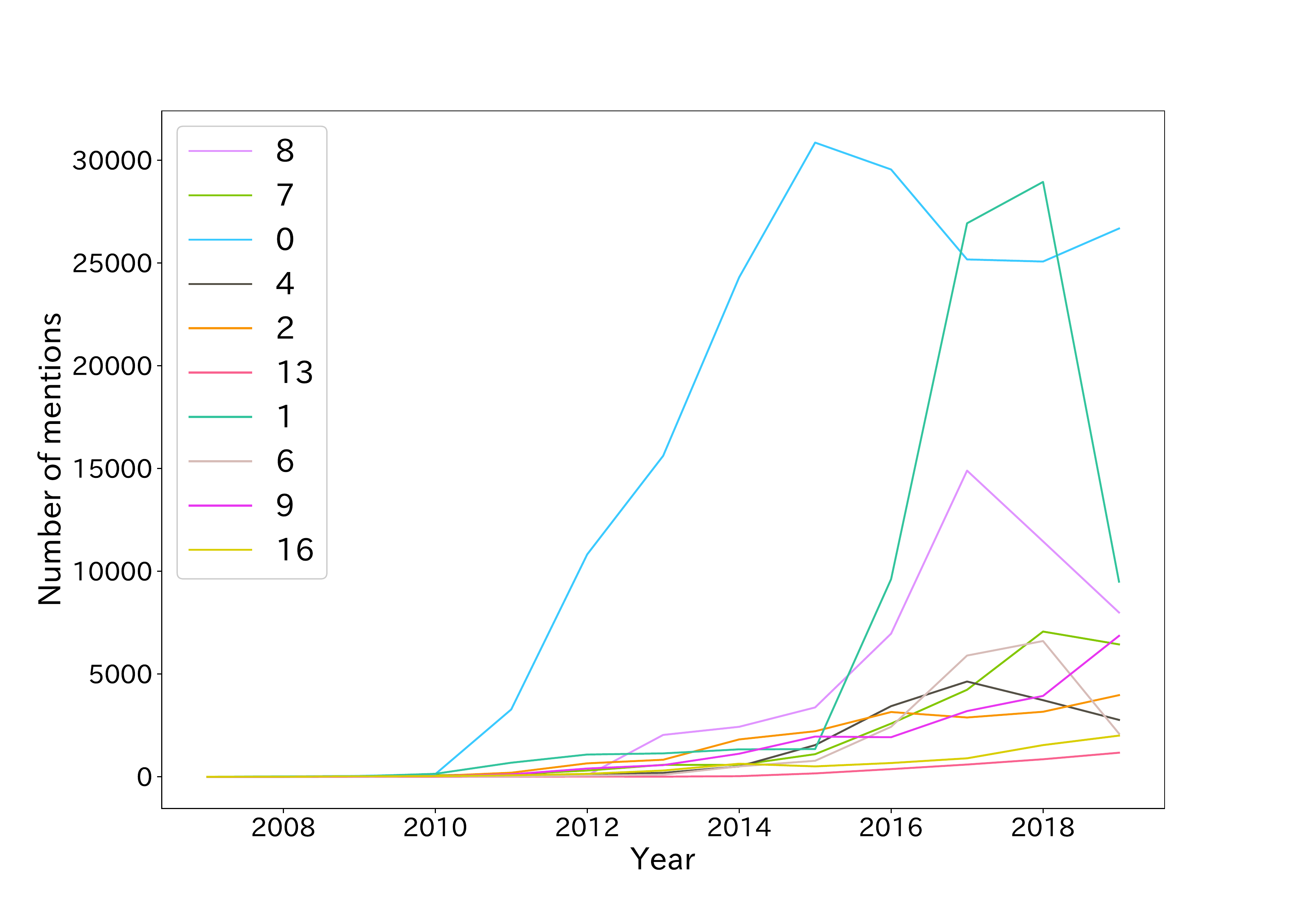}
	\caption{Time-series variation in the number of mentions by community}
	\label{fig:mention_variation}
\end{figure}

However, the mention period of community 13 is particularly short compared with the other communities in Table~\ref{tab:comtime}.
As a result of this observation, we examined the time-series variation in the number of mentions by each community in Figure~\ref{fig:mention_variation}.
The horizontal axis represents the year and the vertical axis represents the number of mentions in the community in that year.
The results indicate that community 13 is a relatively young community, with a rapid increase in mentions since 2016.

In community 13, the most mentioned arXiv paper was about Google's neural machine translation in 2016~\cite{wu2016googles}, and the most liked and retweeted arXiv paper was about Google's BERT in 2019~\cite{devlin2019bert}.
Considering that the main category of community 13 in Table~\ref{tab:comcategory} is natural language processing (cs.CL), we can see that this community is a group of information spreaders who spread information about natural language processing, especially deep learning.
In other words, even in the same technical field of machine learning, different communities were formed due to the difference in the backgrounds of the information spreaders between computer vision (cs.CV) and natural language processing (cs.CL).

\section{Conclusion}
In this paper, we modeled arXiv information distribution by assuming that a user has two types of roles: information spreader and information collector, which enabled us to eliminate bots and focus on users who contribute highly to information distribution on Twitter.
Furthermore, we attempted to analyze the characteristics of users in more detail than previous studies by bringing in two different perspectives, such as authority weight and betweenness centrality, information spread category and information collect category, and communication language and profile language.
From these results, 
we found that information about arXiv papers circulates on Twitter from information spreaders to information collectors and that multiple communities of information spreaders are formed according to their research fields.
It was also found that different communities were formed in the same research field, depending on the research or cultural background of the information spreaders.
We were able to identify two types of key persons: information spreaders who lead the relevant field in the international community and information spreaders who bridge the regional and international communities using English and their native language.
In addition, we found that it takes some time to gain trust as an information spreader.

This analysis was performed on data before the COVID-19 pandemic.
We plan to analyze the new role of arXiv and its changes owing to the current global circumstances.
Furthermore, we plan to study a new bibliographic index based on the analysis of academic information distribution on social media.

\begin{acks}
\begin{sloppypar}
This work was supported by JSPS KAKENHI Grant Number JP19H04421.
\end{sloppypar}
\end{acks}

\bibliographystyle{ACM-Reference-Format}
\bibliography{main}


\begin{thebibliography}{20}

%%% the Web address.

\ifx \showCODEN    \undefined \def \showCODEN     #1{\unskip}     \fi
\ifx \showDOI      \undefined \def \showDOI       #1{#1}\fi
\ifx \showISBNx    \undefined \def \showISBNx     #1{\unskip}     \fi
\ifx \showISBNxiii \undefined \def \showISBNxiii  #1{\unskip}     \fi
\ifx \showISSN     \undefined \def \showISSN      #1{\unskip}     \fi
\ifx \showLCCN     \undefined \def \showLCCN      #1{\unskip}     \fi
\ifx \shownote     \undefined \def \shownote      #1{#1}          \fi
\ifx \showarticletitle \undefined \def \showarticletitle #1{#1}   \fi
\ifx \showURL      \undefined \def \showURL       {\relax}        \fi
% invisible to TeX
\providecommand\bibfield[2]{#2}
\providecommand\bibinfo[2]{#2}
\providecommand\natexlab[1]{#1}
\providecommand\showeprint[2][]{arXiv:#2}

\bibitem[\protect\citeauthoryear{Blondel, Guillaume, Lambiotte, and
  Lefebvre}{Blondel et~al\mbox{.}}{2008}]%
        {Blondel_2008}
\bibfield{author}{\bibinfo{person}{Vincent~D. Blondel},
  \bibinfo{person}{Jean-Loup Guillaume}, \bibinfo{person}{Renaud Lambiotte},
  {and} \bibinfo{person}{Etienne Lefebvre}.} \bibinfo{year}{2008}\natexlab{}.
\newblock \showarticletitle{Fast unfolding of communities in large networks}.
\newblock \bibinfo{journal}{\emph{Journal of Statistical Mechanics: Theory and
  Experiment}} \bibinfo{volume}{2008}, \bibinfo{number}{10}
  (\bibinfo{year}{2008}), \bibinfo{pages}{P10008}.
\newblock


\bibitem[\protect\citeauthoryear{Chiarelli, Johnson, Pinfield, and
  Richens}{Chiarelli et~al\mbox{.}}{2019}]%
        {chiarelli_andrea_2019_3357727}
\bibfield{author}{\bibinfo{person}{Andrea Chiarelli}, \bibinfo{person}{Rob
  Johnson}, \bibinfo{person}{Stephen Pinfield}, {and} \bibinfo{person}{Emma
  Richens}.} \bibinfo{year}{2019}\natexlab{}.
\newblock \bibinfo{title}{Accelerating scholarly communication: The
  transformative role of preprints}.
\newblock
\newblock
\urldef\tempurl%
\url{https://doi.org/10.5281/zenodo.3357727}
\showURL{%
\tempurl}


\bibitem[\protect\citeauthoryear{Devlin, Chang, Lee, and Toutanova}{Devlin
  et~al\mbox{.}}{2019}]%
        {devlin2019bert}
\bibfield{author}{\bibinfo{person}{Jacob Devlin}, \bibinfo{person}{Ming-Wei
  Chang}, \bibinfo{person}{Kenton Lee}, {and} \bibinfo{person}{Kristina
  Toutanova}.} \bibinfo{year}{2019}\natexlab{}.
\newblock \bibinfo{title}{{BERT}: Pre-training of Deep Bidirectional
  Transformers for Language Understanding}.
\newblock
\newblock
\showeprint[arxiv]{1810.04805}~[cs.CL]


\bibitem[\protect\citeauthoryear{Freeman}{Freeman}{1977}]%
        {freeman1977}
\bibfield{author}{\bibinfo{person}{Linton~C. Freeman}.}
  \bibinfo{year}{1977}\natexlab{}.
\newblock \showarticletitle{A set of measures of centrality based on
  betweenness}.
\newblock \bibinfo{journal}{\emph{Sociometry}} \bibinfo{volume}{40},
  \bibinfo{number}{1} (\bibinfo{year}{1977}), \bibinfo{pages}{35--41}.
\newblock


\bibitem[\protect\citeauthoryear{Gentil-Beccot, Mele, and Brooks}{Gentil-Beccot
  et~al\mbox{.}}{2009}]%
        {Gentil-Beccot:1214893}
\bibfield{author}{\bibinfo{person}{Anne Gentil-Beccot},
  \bibinfo{person}{Salvatore Mele}, {and} \bibinfo{person}{Travis~C. Brooks}.}
  \bibinfo{year}{2009}\natexlab{}.
\newblock \showarticletitle{Citing and Reading Behaviours in High-Energy
  Physics: How a Community Stopped Worrying about Journals and Learned to Love
  Repositories}.
\newblock \bibinfo{journal}{\emph{Scientometrics}}  \bibinfo{volume}{84}
  (\bibinfo{year}{2009}), \bibinfo{pages}{345--355. 12 p}.
\newblock


\bibitem[\protect\citeauthoryear{Halpern}{Halpern}{2000}]%
        {DBLP:journals/jcd/Halpern00}
\bibfield{author}{\bibinfo{person}{Joseph~Y. Halpern}.}
  \bibinfo{year}{2000}\natexlab{}.
\newblock \showarticletitle{{CoRR}: a computing research repository}.
\newblock \bibinfo{journal}{\emph{{ACM} Journal of Computer Documentation}}
  \bibinfo{volume}{24}, \bibinfo{number}{2} (\bibinfo{year}{2000}),
  \bibinfo{pages}{41--48}.
\newblock


\bibitem[\protect\citeauthoryear{Ke, Ahn, and Sugimoto}{Ke
  et~al\mbox{.}}{2017}]%
        {10.1371/journal.pone.0175368}
\bibfield{author}{\bibinfo{person}{Qing Ke}, \bibinfo{person}{Yong-Yeol Ahn},
  {and} \bibinfo{person}{Cassidy~R. Sugimoto}.}
  \bibinfo{year}{2017}\natexlab{}.
\newblock \showarticletitle{A systematic identification and analysis of
  scientists on {Twitter}}.
\newblock \bibinfo{journal}{\emph{PLOS ONE}} \bibinfo{volume}{12},
  \bibinfo{number}{4} (\bibinfo{year}{2017}), \bibinfo{pages}{1--17}.
\newblock


\bibitem[\protect\citeauthoryear{Kleinberg}{Kleinberg}{1999}]%
        {Kleinberg:1999:ASH:324133.324140}
\bibfield{author}{\bibinfo{person}{Jon~M. Kleinberg}.}
  \bibinfo{year}{1999}\natexlab{}.
\newblock \showarticletitle{Authoritative Sources in a Hyperlinked
  Environment}.
\newblock \bibinfo{journal}{\emph{J. ACM}} \bibinfo{volume}{46},
  \bibinfo{number}{5} (\bibinfo{year}{1999}), \bibinfo{pages}{604--632}.
\newblock


\bibitem[\protect\citeauthoryear{Larivière, Sugimoto, Macaluso, Milojević,
  Cronin, and Thelwall}{Larivière et~al\mbox{.}}{2014}]%
        {doi:10.1002/asi.23044arXivToDB}
\bibfield{author}{\bibinfo{person}{Vincent Larivière},
  \bibinfo{person}{Cassidy~R. Sugimoto}, \bibinfo{person}{Benoit Macaluso},
  \bibinfo{person}{Staša Milojević}, \bibinfo{person}{Blaise Cronin}, {and}
  \bibinfo{person}{Mike Thelwall}.} \bibinfo{year}{2014}\natexlab{}.
\newblock \showarticletitle{{arXiv} E-prints and the journal of record: An
  analysis of roles and relationships}.
\newblock \bibinfo{journal}{\emph{Journal of the Association for Information
  Science and Technology}} \bibinfo{volume}{65}, \bibinfo{number}{6}
  (\bibinfo{year}{2014}), \bibinfo{pages}{1157--1169}.
\newblock


\bibitem[\protect\citeauthoryear{Lin, Yu, Zhou, Zhou, and Shi}{Lin
  et~al\mbox{.}}{2020}]%
        {howmanyPreprint}
\bibfield{author}{\bibinfo{person}{Jialiang Lin}, \bibinfo{person}{Yao Yu},
  \bibinfo{person}{Yu Zhou}, \bibinfo{person}{Zhiyang Zhou}, {and}
  \bibinfo{person}{Xiaodong Shi}.} \bibinfo{year}{2020}\natexlab{}.
\newblock \showarticletitle{How many preprints have actually been printed and
  why: a case study of computer science preprints on {arXiv}}.
\newblock \bibinfo{journal}{\emph{Scientometrics}}  \bibinfo{volume}{124}
  (\bibinfo{year}{2020}), \bibinfo{pages}{1--20}.
\newblock


\bibitem[\protect\citeauthoryear{Mohammadi, Thelwall, Kwasny, and
  Holmes}{Mohammadi et~al\mbox{.}}{2018}]%
        {10.1371/journal.pone.0197265}
\bibfield{author}{\bibinfo{person}{Ehsan Mohammadi}, \bibinfo{person}{Mike
  Thelwall}, \bibinfo{person}{Mary Kwasny}, {and} \bibinfo{person}{Kristi~L.
  Holmes}.} \bibinfo{year}{2018}\natexlab{}.
\newblock \showarticletitle{Academic information on {Twitter}: A user survey}.
\newblock \bibinfo{journal}{\emph{PLOS ONE}} \bibinfo{volume}{13},
  \bibinfo{number}{5} (\bibinfo{year}{2018}), \bibinfo{pages}{1--18}.
\newblock


\bibitem[\protect\citeauthoryear{Rieger, Steinhart, and Cooper}{Rieger
  et~al\mbox{.}}{2016}]%
        {rieger2016arxiv25}
\bibfield{author}{\bibinfo{person}{Oya~Y. Rieger}, \bibinfo{person}{Gail
  Steinhart}, {and} \bibinfo{person}{Deborah Cooper}.}
  \bibinfo{year}{2016}\natexlab{}.
\newblock \bibinfo{title}{{arXiv@25}: Key findings of a user survey}.
\newblock
\newblock
\showeprint[arxiv]{1607.08212}~[cs.DL]


\bibitem[\protect\citeauthoryear{Shen, Ma, and Wang}{Shen
  et~al\mbox{.}}{2018}]%
        {shen-etal-2018-web}
\bibfield{author}{\bibinfo{person}{Zhihong Shen}, \bibinfo{person}{Hao Ma},
  {and} \bibinfo{person}{Kuansan Wang}.} \bibinfo{year}{2018}\natexlab{}.
\newblock \showarticletitle{A Web-scale system for scientific knowledge
  exploration}. In \bibinfo{booktitle}{\emph{Proceedings of {ACL} 2018, System
  Demonstrations}}. \bibinfo{pages}{87--92}.
\newblock


\bibitem[\protect\citeauthoryear{Shuai, Pepe, and Bollen}{Shuai
  et~al\mbox{.}}{2012}]%
        {10.1371/journal.pone.0047523OnlineReaction}
\bibfield{author}{\bibinfo{person}{Xin Shuai}, \bibinfo{person}{Alberto Pepe},
  {and} \bibinfo{person}{Johan Bollen}.} \bibinfo{year}{2012}\natexlab{}.
\newblock \showarticletitle{How the Scientific Community Reacts to Newly
  Submitted Preprints: Article Downloads, {Twitter} Mentions, and Citations}.
\newblock \bibinfo{journal}{\emph{PLOS ONE}} \bibinfo{volume}{7},
  \bibinfo{number}{11} (\bibinfo{year}{2012}), \bibinfo{pages}{1--8}.
\newblock


\bibitem[\protect\citeauthoryear{Sinha, Shen, Song, Ma, Eide, Hsu, and
  Wang}{Sinha et~al\mbox{.}}{2015}]%
        {mag2015}
\bibfield{author}{\bibinfo{person}{Arnab Sinha}, \bibinfo{person}{Zhihong
  Shen}, \bibinfo{person}{Yang Song}, \bibinfo{person}{Hao Ma},
  \bibinfo{person}{Darrin Eide}, \bibinfo{person}{Bo-June~(Paul) Hsu}, {and}
  \bibinfo{person}{Kuansan Wang}.} \bibinfo{year}{2015}\natexlab{}.
\newblock \showarticletitle{An Overview of Microsoft Academic Service ({MAS})
  and Applications}. In \bibinfo{booktitle}{\emph{Proceedings of the 24th
  International Conference on World Wide Web}}. \bibinfo{pages}{243–246}.
\newblock


\bibitem[\protect\citeauthoryear{Sutton and Gong}{Sutton and Gong}{2017}]%
        {sutton2017popularity}
\bibfield{author}{\bibinfo{person}{Charles Sutton} {and} \bibinfo{person}{Linan
  Gong}.} \bibinfo{year}{2017}\natexlab{}.
\newblock \bibinfo{title}{Popularity of {arXiv.org} within Computer Science}.
\newblock
\newblock
\showeprint[arxiv]{1710.05225}~[cs.DL]


\bibitem[\protect\citeauthoryear{Vrettas and Sanderson}{Vrettas and
  Sanderson}{2015}]%
        {ConferencesVS}
\bibfield{author}{\bibinfo{person}{George Vrettas} {and} \bibinfo{person}{Mark
  Sanderson}.} \bibinfo{year}{2015}\natexlab{}.
\newblock \showarticletitle{Conferences versus journals in computer science}.
\newblock \bibinfo{journal}{\emph{Journal of the Association for Information
  Science and Technology}} \bibinfo{volume}{66}, \bibinfo{number}{12}
  (\bibinfo{year}{2015}), \bibinfo{pages}{2674--2684}.
\newblock


\bibitem[\protect\citeauthoryear{Wang, Chen, and Glänzel}{Wang
  et~al\mbox{.}}{2020}]%
        {WANG2020101097}
\bibfield{author}{\bibinfo{person}{Zhiqi Wang}, \bibinfo{person}{Yue Chen},
  {and} \bibinfo{person}{Wolfgang Glänzel}.} \bibinfo{year}{2020}\natexlab{}.
\newblock \showarticletitle{Preprints as accelerator of scholarly
  communication: An empirical analysis in Mathematics}.
\newblock \bibinfo{journal}{\emph{Journal of Informetrics}}
  \bibinfo{volume}{14}, \bibinfo{number}{4} (\bibinfo{year}{2020}),
  \bibinfo{pages}{101097}.
\newblock


\bibitem[\protect\citeauthoryear{Wu, Schuster, Chen, Le, Norouzi, Macherey,
  Krikun, Cao, Gao, Macherey, Klingner, Shah, Johnson, Liu, Łukasz Kaiser,
  Gouws, Kato, Kudo, Kazawa, Stevens, Kurian, Patil, Wang, Young, Smith, Riesa,
  Rudnick, Vinyals, Corrado, Hughes, and Dean}{Wu et~al\mbox{.}}{2016}]%
        {wu2016googles}
\bibfield{author}{\bibinfo{person}{Yonghui Wu}, \bibinfo{person}{Mike
  Schuster}, \bibinfo{person}{Zhifeng Chen}, \bibinfo{person}{Quoc~V. Le},
  \bibinfo{person}{Mohammad Norouzi}, \bibinfo{person}{Wolfgang Macherey},
  \bibinfo{person}{Maxim Krikun}, \bibinfo{person}{Yuan Cao},
  \bibinfo{person}{Qin Gao}, \bibinfo{person}{Klaus Macherey},
  \bibinfo{person}{Jeff Klingner}, \bibinfo{person}{Apurva Shah},
  \bibinfo{person}{Melvin Johnson}, \bibinfo{person}{Xiaobing Liu},
  \bibinfo{person}{Łukasz Kaiser}, \bibinfo{person}{Stephan Gouws},
  \bibinfo{person}{Yoshikiyo Kato}, \bibinfo{person}{Taku Kudo},
  \bibinfo{person}{Hideto Kazawa}, \bibinfo{person}{Keith Stevens},
  \bibinfo{person}{George Kurian}, \bibinfo{person}{Nishant Patil},
  \bibinfo{person}{Wei Wang}, \bibinfo{person}{Cliff Young},
  \bibinfo{person}{Jason Smith}, \bibinfo{person}{Jason Riesa},
  \bibinfo{person}{Alex Rudnick}, \bibinfo{person}{Oriol Vinyals},
  \bibinfo{person}{Greg Corrado}, \bibinfo{person}{Macduff Hughes}, {and}
  \bibinfo{person}{Jeffrey Dean}.} \bibinfo{year}{2016}\natexlab{}.
\newblock \bibinfo{title}{Google's Neural Machine Translation System: Bridging
  the Gap between Human and Machine Translation}.
\newblock
\newblock
\showeprint[arxiv]{1609.08144}~[cs.CL]


\bibitem[\protect\citeauthoryear{Xie, Shen, and Wang}{Xie
  et~al\mbox{.}}{2021}]%
        {mspreprint2021}
\bibfield{author}{\bibinfo{person}{Boya Xie}, \bibinfo{person}{Zhihong Shen},
  {and} \bibinfo{person}{Kuansan Wang}.} \bibinfo{year}{2021}\natexlab{}.
\newblock \bibinfo{title}{Is preprint the future of science? {A} thirty year
  journey of online preprint services}.
\newblock
\newblock
\showeprint[arxiv]{2102.09066}~[cs.DL]


\end{thebibliography}
\end{document}